\documentclass[%
 preprint,
 amsmath,amssymb,
 aps,
]{revtex4-2}

\usepackage{graphicx}
\usepackage{dcolumn}
\usepackage{bm}
\usepackage{comment}
\usepackage{xcolor}
\usepackage{float}

\def\yb{\mathbf{y}}
\def\xb{\mathbf{x}}
\def\etab{\boldsymbol{\eta}}
\def\xib{\boldsymbol{\xi}}

\def\xtb{\tilde{\mathbf{x}}}
\def\etatb{\tilde{\boldsymbol{\eta}}}

\DeclareMathOperator{\csch}{csch}

\begin{document}

\preprint{APS/123-QED}

\title{Most probable paths for active Ornstein-Uhlenbeck particles}
\thanks{A footnote to the article title}%

\author{Sandipan Dutta}
 \email{sandipan.dutta@pilani.bits-pilani.ac.in}
\affiliation{%
 Department of Physics, Birla Institute of Technology and Science, Pilani, Rajasthan, 333031, India\\
}%

\date{\today}

\begin{abstract}
Fluctuations play an important role in the dynamics of stochastic systems. In particular, for small systems, the most probable thermodynamic quantities differ from their averages because of the fluctuations. Using the Onsager Machlup variational formalism we analyze the most probable paths for non-equilibrium systems, in particular active Ornstein-Uhlenbeck particles (AOUP), and investigate how the entropy production along these paths differ from the average entropy production. We investigate how much information about their non-equilibrium nature can be obtained from their extremum paths and how these paths depend on the persistence time and their swim velocities. We also look at how the entropy production along the most probable paths varies with the active noise and how it differs from the average entropy production. This study would be useful to design artificial active systems with certain target trajectories.  
\end{abstract}

\maketitle


\section{Introduction}

Active matter is a new class of nonequilibrium systems in condensed matter which constantly dissipates energy to produce motion \cite{marchetti2013hydrodynamics, bechinger2016active,  elgeti2015physics,  vicsek2012collective}. Flocking of birds and fishes and swimming of organisms are examples of active matter systems \cite{ramaswamy2010mechanics}. Many artificial motile systems like Janus colloids \cite{howse2007self, jiang2010active, volpe2011microswimmers}, colloidal rollers \cite{bricard2015emergent} and water droplets \cite{izri2014self, thutupalli2011swarming} have been developed that mimics these biological systems. Several models have been used to study active systems like the active Brownian particle \cite{romanczuk2012active, ten2011brownian, sevilla2015smoluchowski}, run and tumble \cite{nash2010run, tailleur2008statistical, angelani2015run} and active Ornstein–Uhlenbeck particle (AOUP) model \cite{szamel2014self, maggi2015marini, shaebani2020computational, caprini2022parental}. 
AOUP is one of the simplest models where the velocities of the active particles are exponentially correlated in time and is modeled using the Langevin equations with colored noise.  All these models have been able to explain the collective behavior of active systems called motility -induced phase separation\cite{cates2015annrev, liebchen2018synthetic,tailleur2008statistical, stenhammar2013continuum} through a combination of motility and steric repulsion. While most studies on active matter look at their collective behavior, here we focus on the thermodynamics along the trajectories of a single active particle.

In many of these systems, the correlated noise like the Ornstein-Uhlenbeck (OU) noise induces stochastic transitions between possible states, as in the case of chemical reactions \cite{hanggi1990reaction, tripathi2022acceleration}. Many of these transitions in the two state systems are induced by noise. These stochastic transitions are rare with a very low probability like a microswimmer passing through a slit \cite{salek2019bacterial} and escaping from a capture near a wall \cite{elgeti2009self}. These rare events help microorganisms in their survival. Finding the most probable trajectory between two given points is one of the key problems in the rare events \cite{durr1978onsager, wissel1979manifolds, faccioli2006dominant, adib2008stochastic, wang2010kinetic}. The path probability and the Onsager-Machlup (OM) integral are useful tools to calculate the most probable path (MPP) for
arbitrary initial and final states \cite{onsager1953fluctuations, machlup1953fluctuations, doi2019application, wang2021onsager, cates2022stochastic}. The MPP of transitions have been studied in a double-well potential \cite{adib2008stochastic}, experiments \cite{gladrow2021experimental}, protein folding \cite{faccioli2006dominant, yasuda2022onsager} and chemical reactions \cite{wang2010kinetic}. Interest in the MPPs in active systems is very recent \cite{yasuda2022most}. Yasuda and Ishimoto recently showed that the extremum path of a single active Brownian particle is analogous to the pendulum equation. They obtained multiple extemum paths of which they found the U-shaped path to be most probable. 

In this work, we find the MPP for a system of AOUP particles from the extremum of the OM integral. We focus on the case of non-interacting AOUP particles for which the MPPs are analytically solvable for all boundary conditions unlike ABP particles \cite{yasuda2022most} and obtain their phase diagram. Our primary objective is to investigate how much information about the non-equilibrium nature of the AOUP is contained in their extremum trajectories. We look at how the trajectories differ at different regions of the phase space. We find that the AOUP has an unstable fixed point and study its dynamics around that fixed point. We then consider the entropy production for the MPP. The entropy production quantifies the time irreversibility of the non-equilibrium systems \cite{dabelow2019irreversibility} and is obtained from the ratio of the probabilities of the forward and backward trajectories \cite{lebowitz1999gallavotti}. It has been suggested \cite{dabelow2019irreversibility} that the OU noise can have different parities under time reversal in the context of active particles in a thermal bath or second, a passive particle in an active bath. The OU noise is odd under time-reversal \cite{dabelow2019irreversibility} for an active AOUP particle in an equilibrium thermal bath \cite{maggi2014generalized, maggi2017memory, argun2016non, chaki2018entropy} and even for a passive particle in an active AOUP bath \cite{fily2012, maggi2015marini, paoluzzi2016critical, farage2015effective, nardini2017entropy, marconi2017heat, mandal2018mandal, flenner2016nonequilibrium, berthier2013non, shankar2018hidden}.  Accordingly the entropy productions are different in these cases. Here we consider both the parities and compare the differences in behavior. We also compare the entropy production for the MPP with the average entropy production.    

In the next section, we write the stochastic dynamics of AOUP particles. The MPP is derived from the OM integral for the AOUP system in Section \ref{sec: most probable path} and the phase space of solutions is obtained in Section \ref{sec:freeaoup} for the case of a single AOUP particle. We calculate the entropy production along the most probable and extremum trajectories in Section \ref{sec:entropy} and compare it with the average entropy production. Finally we conclude in Section \ref{sec:conclusion}.   

\section{active Ornstein–Uhlenbeck particle (AOUP) model}
\label{sec: aoup}
Consider an active particle suspended in a thermal bath of temperature $T$. The particles experience a conservative force $ -\nabla U(\mathbf{x})$ in addition to active forces $\boldsymbol{\eta}$ from their intrinsic self-propulsion. $U$ is the potential energy of a conservative force due to the interactions among the particles as well as some external potential, and $\mathbf{x}$ is the position of the particle. This description applies to both cases of a passive particle in an active bath or an active particle in a passive bath. 

We consider the overdamped Langevin equation \cite{van1992stochastic, gardiner1985handbook} for the particle
\begin{equation}
\dot{\xb}(t) = \mathbf{f}(\xb(t),t) + \sqrt{2D_a}\etab(t) + \sqrt{2D}\xib(t),
\label{langevin}
\end{equation}
where $\mathbf{f}(\xb,t) = -\frac{1}{\gamma}\nabla U(\xb,t)$ and $\gamma$ is the hydrodynamic friction. $\xib(t)$ are the thermal fluctuations which are Gaussian with zero mean and delta function correlations $\langle\xi_{\alpha}(t)\xi_{\beta}(t^{\prime})\rangle = \delta_{\alpha\beta}\delta(t-t^{\prime})$, where $\alpha$ and $\beta$ are the Cartesian indices. $D = k_BT/\gamma$ is the diffusion constant in the thermal bath and $D_a$ is the diffusion constant due to the activity. The active fluctuations $\etab$ are OU processes with zero mean and exponentially correlations
\begin{equation}
\langle\eta_{\alpha}(t)\eta_{\alpha}(t^{\prime})\rangle = \frac{\delta_{\alpha\beta}}{2\tau}\exp(-\vert t-t^{\prime}\vert/\tau).
\label{ounoise}
\end{equation}
$\tau$ is the persistence time that quantifies the persistence of the active fluctuations.  The active diffusion constant can be written as $D_a = v_0^2\tau$ where $v_0$ is the swim velocity of the active particles. \cite{caprini2021correlated} These equations are easily generalized to $N$ identical particles with positions $\{\xb_i\}$ with the same parameters $D$ and $D_a$. The active noise for two particles satisfies $\langle\eta_{i\alpha}\eta_{j\beta}\rangle = \frac{\delta_{\alpha\beta}\delta_{ij}}{2\tau}\exp(-\vert t-t^{\prime}\vert/\tau)$, where $i$, $j$ are particle indices.

\section{Most probable trajectory}
\label{sec: most probable path}
Let $\yb = (\xb,\etab)$ denote the combined variable of the particle position and the active noise. The probability of observing certain trajectory of the Markovian process $\yb(s)\vert_0^t$ starting at $\yb_0$ at $s = 0$ and ending at $\yb_f$ at $s = t$ is given by the Onsager-Machlup path integral \cite{onsager1953fluctuations, machlup1953fluctuations} by
\begin{equation}
p(\yb(t)\vert\yb_0)  \propto \exp\biggl[-\int_0^t\frac{ds}{4}\sum_{i,j}\sum_{\alpha,\beta}\left(f_{i\alpha}(\yb(s))-\dot{y}_{i\alpha}\right)D^{-1}_{i\alpha, j\beta}\left(f_{j\beta}(\yb(s))-\dot{y}_{j\beta}\right)\biggr].
\label{onsagermaclup}
\end{equation}
$y_{i\alpha}$ denotes the $\alpha$ component of  variable $\yb$ for the particle $i$. $D_{i\alpha, j\beta}$ is the diffusion tensor that is diagonal in our case. This equation is derived in the Appendix \ref{appendix: onsagermaclup} following Reference \cite{risken1984solutions, yasuda2022most}. For the Langevin equation \eqref{langevin} summing over $\alpha$, $\beta$ and putting explicit form of $\yb$ in terms of $\xb$ and $\etab$ and the diffusion matrix (Appendix \ref{appendix: onsagermaclup}),  we get \cite{dabelow2019irreversibility, dabelow2021irreversible}
\begin{equation}
p(\xb,\etab\vert\xb_0,\etab_0)  \propto\exp\biggl(-\int_0^tds\sum_{i=1}^N\biggl[\frac{(\dot{\xb}_i(s)-\mathbf{f}_i(s)-\sqrt{2D_a}\etab_i(s))^2}{4D}+\frac{(\tau\dot{\etab}_i(s)+\etab_i(s))^2}{2}+\frac{1}{2}\nabla.\mathbf{f}_i(s)\biggr]\biggr).
\label{eichhorn}
\end{equation}
The last term in the exponent occurs from the indeterminacy of the time discretization \cite{yasuda2022most}.

The quantity inside the exponent of the OM integral is similar to a Lagrangian and we denote it by   
\begin{equation}
L(\xb,\dot{\xb},\etab,\dot{\etab}) = \sum_{i=1}^N\frac{(\dot{\xb}_i(s)-\mathbf{f}_i(s)-\sqrt{2D_a}\etab_i(s))^2}{4D}+\frac{(\tau\dot{\etab}_i(s)+\etab_i(s))^2}{2}+\frac{1}{2}\nabla.\mathbf{f}_i(s). 
\label{lagrangian}
\end{equation}
The extremum paths are obtained by the variation of the OM integral and setting the first variation to zero 
\begin{equation}
\frac{d}{dt}\frac{\delta}{\delta \dot{y}_{i\alpha}(t)}L(\mathbf{x},\dot{\mathbf{x}})-\frac{\delta}{\delta y_{i\alpha}(t)}L(\mathbf{x},\dot{\mathbf{x}})=0,
\end{equation}
which from Equation \eqref{lagrangian} gives 
\begin{align}
\frac{1}{2D}(\ddot{x}_{i\alpha}-\frac{d}{dt}f_{i\alpha}-\sqrt{2D_a}\dot{\eta}_{i\alpha}) & = -\frac{1}{2D}\partial_{i\alpha}f_{j\beta}(\dot{x}_{j\beta}-f_{j\beta}-\sqrt{2D_a}\eta_{j\beta})+\frac{1}{2}\partial_{i\alpha}\partial_{j\beta}f_{j\beta},\nonumber\\
(\tau^2\ddot{\eta}_{i\alpha}+\tau\dot{\eta}_{i\alpha}) & = (\tau\dot{\eta}_{i\alpha}+\eta_{i\alpha})-\frac{\sqrt{2D_a}}{2D}(\dot{x}_{i\alpha}-f_{i\alpha}-\sqrt{2D_a}\eta_{i\alpha}).
\label{mostprobablepath}
\end{align}
These coupled equations are solved to obtain the extremum paths. The second variation of the OM integral should be positive for the probability to be maximum along the extremum path.

\section{Free active AOUP particles}
\label{sec:freeaoup}
The MPP in Equation \eqref{mostprobablepath} can be solved analytically in the case of free AOUP particles for which $\mathbf{f} = 0$. The equations of motion of the particles decouple from one another because of the absence of the interactions. The motion along the $x$, $y$ and $z$ coordinates also decouples due to the nature of OU noise. Thus for the free AOUP system we will focus on the dynamics of a single particle in one dimension. 

\subsection{The most probable path in one dimension}
\label{sec: mostprobablepathfreeaoup}
The extremum path for a single AOUP particle in one dimension is given by
\begin{align}
\ddot{x}-\sqrt{2D_a}\dot{\eta} & = 0\nonumber\\
\tau^2\ddot{\eta} & = \eta-\frac{\sqrt{2D_a}}{2D}(\dot{x}-\sqrt{2D_a}\eta)
\label{eq:extremumpath}
\end{align}

\begin{figure}[!htb]
\includegraphics[width=1\textwidth]{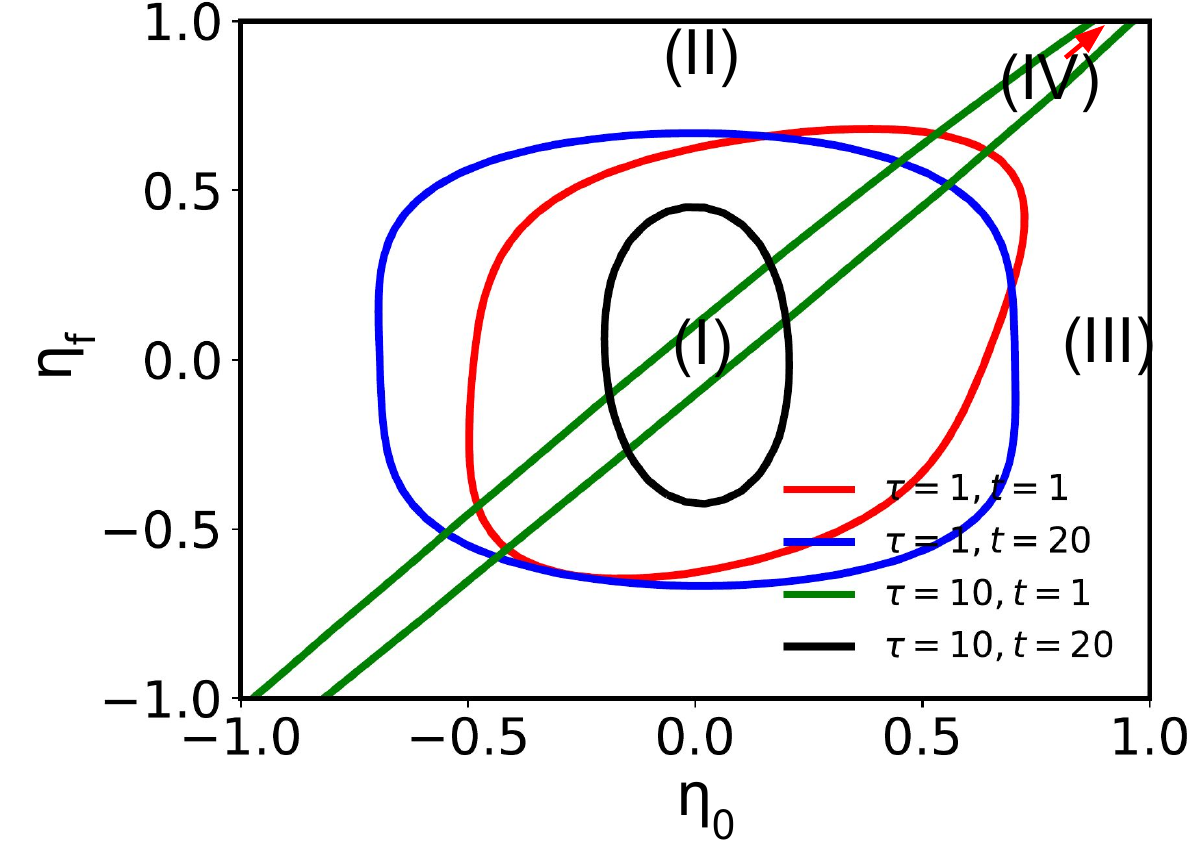}
\caption{ The constant $L = 0.5$ curves for different initial and final active noise parameters $\eta_0$ and $\eta_f$. The paths that satisfies Eq \eqref{eq:extremumpath} and $L\approx 0$ would give the MPPs. $t$ is the travel time for the particle to reach $x(t) = 1$ from the origin $x(0) = 0$ and persistence times $\tau$. The other parameters are fixed at $D = 1$ and $D_a = 5$. We plot the trajectories at the points $\eta_0 = 0, \eta_f = 0$ (I), $\eta_0 = 0, \eta_f = 1$ (II), $\eta_0 = 1, \eta_f = 0$ (III) and $\eta_0 = 5, \eta_f = 5$ (IV) in Fig \ref{fig:mpp}.  }
\label{fig:solution_space}
\end{figure}

Since the whole derivation is based on the use of translational probability in Eq \eqref{onsagermaclup}, we only solve for the trajectories with a fixed start $x(0) = x_0$ and end point $x(t) = x_f$ for a travel time $t$. The active noise also has fixed boundaries $\eta(0) = \eta_0$ and $\eta(t) = \eta_f$. The solution gives the extremum trajectory of the particle which is nothing but the local minimum of the function L in Eq \eqref{eq:extremumpath}. For the rest of the discussions we explore the behavior of the extremum paths of the particle that always starts at the origin $x_0 = 0$ and ends at $x_f = 1$. The exact analytical solutions are obtained using MATHEMATICA software \cite{Mathematica} for $x(t)$ and $\eta(t)$ are given in Appendix \ref{app:solution}.

\begin{figure*}[!htb]
\includegraphics[width=\textwidth]{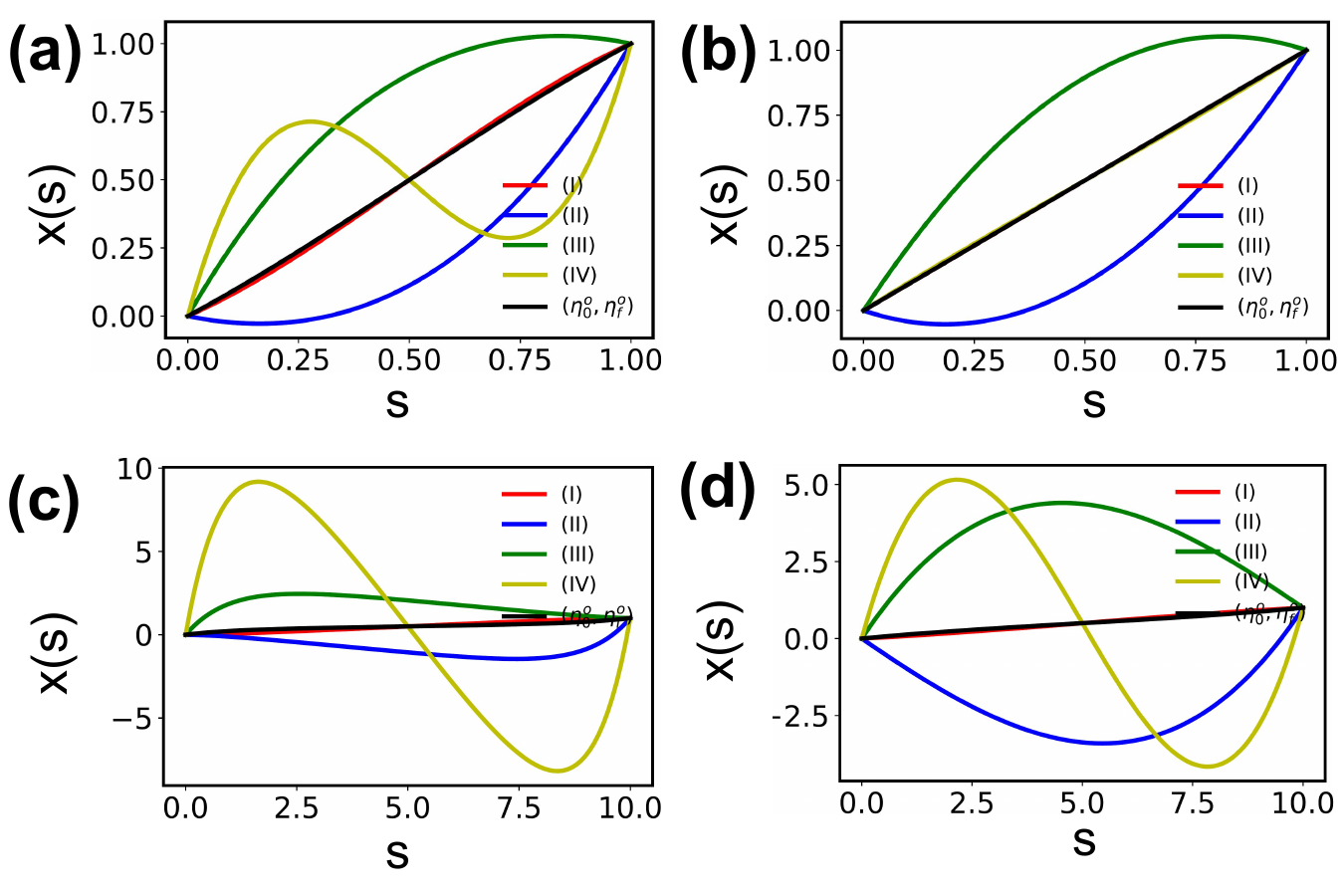}
\caption{The extremum paths of the AOUP between $x(0)=0$ and $x(t)= 1$ at the points (I), (II), (III) and (IV) in Fig \ref{fig:solution_space} for short time $t = 1$ : (a) $\tau = 1$, (b) $\tau = 10$, and long time $t = 10$ : (c) $\tau = 1$, and (d) $\tau= 10$. Also shown are the MPPs for parameters $(\eta_0 = \eta_0^{o}, \eta_f = \eta_f^{o})$ which $L$ function is minimum.}
\label{fig:mpp}
\end{figure*}

\begin{figure*}[!htb]
\includegraphics[width=\textwidth]{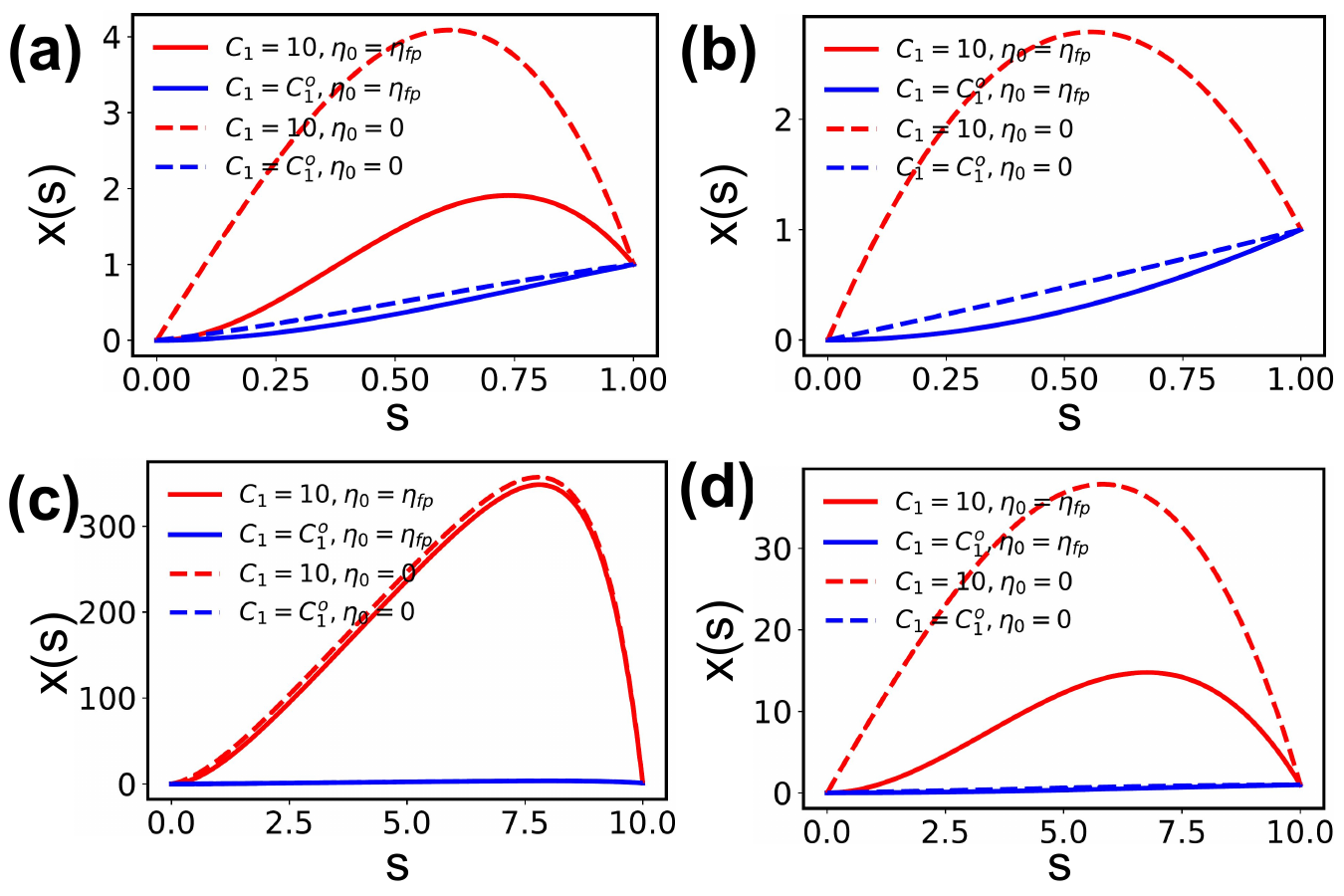}
\caption{The extremum path for $C_1 = 10$ and the MPP of the AOUP near the fixed point $\eta_{fp}$ of Eq \eqref{eq:extremumpathintegrate}  for  (a) $t = 1, \tau =1$, (b) $t = 1, \tau = 10$, (c) $t = 10, \tau = 1$, and (d) $t = 1, \tau = 10$.  }
\label{fig:fixedpoint}
\end{figure*}

These solutions are the local minimum of the $L$ function. To obtain the global minimum of $L$ for the MPPs, we search the phase space of the parameters $\eta_0$ and $\eta_f$ keeping other parameters $D_a$, $D$ and $\tau$ fixed. Since $L\ge 0$ for a free AOUP particle, the global minimum is expected to be $L \approx 0$. From now on we fix the diffusion constants $D = 1$ and $D_a = 5$. The space of solutions of the MPP for which $L\approx 0$ lies inside of the closed curves of Fig \ref{fig:solution_space}. As we increase the persistence time $\tau$, the solution space of the MPP shifts close to the origin and the area inside the curve decreases. This implies that the MPP at large $\tau$ is obtained for very small active noise parameters $\eta_0\approx 0$ and $\eta_f\approx 0$. 

In Fig \ref{fig:mpp} we look at the trajectories at different regions of the $(\eta_0,\eta_f)$ space, in particular at the four points (I), (II), (III) and (IV) in Fig \ref{fig:solution_space}. We numerically obtain the optimal noise parameters $(\eta_0^{(o)},\eta_f^{(o))})$ that gives the smallest values of the $L$ function for short trajectory time $t = 1$ in Fig \ref{fig:mpp}-(a) - (b) and for long time $t = 20$ in Fig \ref{fig:mpp}-(c) - (d). For points (I) and (IV) along $\eta_0 = \eta_f$ line in Fig \ref{fig:solution_space}, the trajectories oscillate about the straight line joining the end points $x(0) = 0$ and $x(t) = 1$ and cross at $t/2$. For $\eta_0 > \eta_f$ (III) the trajectories are above the straight line while for $\eta_0 < \eta_f$ (II) they are stay below. $\eta_0 = 0, \eta_f = 0$ is always a good approximation for the MPP. Larger $L$ values result in longer trajectories and deviation from the straight line or the MPP. For low $\tau$ the AOUP has larger swim velocity thus reaching the end point faster than for large $\tau$. As seen from Fig \ref{fig:solution_space} at large $\tau$, the optimal noise becomes zero thus slowing the particle down. The most probable trajectory of a passive particle between two points $x_0$ and $x_f$ is obtained by taking $D_a \rightarrow 0$ limit in Eq \eqref{eq:trajectory}: 
\begin{equation}
x(s) = (1-\frac{s}{t})x_0+\frac{s}{t}x_f.
\label{eq:passive}
\end{equation}
In large $\tau$ limit Eq \eqref{eq:trajectory} reads: 
\begin{equation}
x(s) = (1-\frac{s}{t})x_0+\frac{s}{t}x_f +\sqrt{2D_a}\frac{s}{2}\left(\frac{s}{t}-1\right)(\eta^{(o)}_f-\eta^{(o)}_0). 
\label{eq:largetau}
\end{equation}
Since $\eta^{(o)}_0\approx 0$ and $\eta^{(o)}_f\approx 0$ from Fig \ref{fig:solution_space}, MPP Eq \eqref{eq:largetau} at large $\tau$ coincides with the MPP of the passive particle in Eq \eqref{eq:passive}. This is also observed in all the plots in Fig \ref{fig:mpp}, as $\tau$ increases it coincides with the straight line joining $x_0$ and $x_f$. This can be understood from the fact that as the persistence time $\tau$ increases for a fixed $D_a$, the swim velocity $v_0 = \sqrt{D_a/\tau}$ decreases. Thus the AOUP behaves like passive particle at large $\tau$.  While for large $t$ (Fig \ref{fig:mpp}-(d)) even a small perturbation of the noise from the optimal noise produce a large deviation from the MPP, for short time $t = 1$ (Fig \ref{fig:mpp}-(b)) the particle does not get enough time to deviate appreciatively from the MPP. 

 We analyze the fixed point dynamics of the AOUP. We rewrite Eq \eqref{eq:extremumpath} as
\begin{align}
 & v_{x}-\sqrt{2D_a}\eta = C_1 \nonumber\\
 & \tau^2 v_{\eta}\frac{d v_{\eta}}{d\eta}=\eta-\frac{\sqrt{2D_a}}{2D}C_1,
\label{eq:extremumpathintegrate}
\end{align}
where $v_x = \dot{x}$, $v_{\eta}=\dot{\eta}$ and $C_1$ is a constant of integration. $C_1$ is the velocity of the particle in absence of the active noise. The drift of the AOUP particle is caused by the active noise and has an unstable fixed point at $\eta_{fp} = -C_1/\sqrt{2D_a}$. The dynamics of the active noise evolves independent of the particle position or velocity because of translational invariance. The second equation of Eq \eqref{eq:extremumpathintegrate} is integrated to obtain
\begin{equation}
v_{\eta}^2 = V_0^2+\frac{1}{\tau^2}\left(\eta-\frac{\sqrt{2}D_a}{2D}C_1\right)^2.
\label{eq:fixedpoints}
\end{equation}
$V_0$ is the velocity at $\eta = \frac{\sqrt{2}D_a}{2D}C_1$. In Fig.\ref{fig:fixedpoint} we see that the extremum trajectories passing through the fixed point $\eta = \eta_{fp}$ are less spread out as the active noise and hence the velocity of the AOUP is smaller. The MPPs for all $t$ and $\tau$ are always close to the passive trajectory of the straight line \ref{eq:passive}. When $t\approx\tau$, the trajectories have similar behavior in Fig \ref{fig:fixedpoint}-(a) and (d) except the trajectories are more spread out in Fig \ref{fig:fixedpoint}-(d) where $t$ is large. For smaller $\tau$ and large $t$ in Fig \ref{fig:fixedpoint}-(c) the swim velocity is large and especially for large $C_1$ the trajectories spread out significantly.

\section{Entropy production for AOUP particles}
\label{sec:entropy}

The non-equilibrium nature of active systems is reflected in the time irreversibility of their trajectories. This is quantified by the entropy production which is the ratio of the probabilities of the forward and the backward paths \cite{lebowitz1999gallavotti}
\begin{equation}
\exp(\Sigma) = \frac{p(\xb,\etab\vert\xb_0,\etab_0)}{\tilde{p}(\xtb,\etatb\vert\xtb_0,\etatb_0)}.
\label{eq:entropyproduction}
\end{equation}
 In order to obtain the time reversed trajectory for a time interval $t$, we are left with the choice of the noise $\etab$ being even or odd under time reversal $\etatb_{\pm}(s) = \pm\etab(t-s)$ . Ref \cite{dabelow2019irreversibility} interpreted the active fluctuations as external forces to have even parity for passive particles in an active bath, while having odd parity for the case of self-propelled particles representing the velocity of the particle. This interpretation of parity has been contested in Ref \cite{caprini2018comment}. The authors in Ref \cite{caprini2019entropy} obtained the entropy production without explicit assumptions about the parity under time reversal. Here we consider both the parities for the active noise and analyze the time reversed trajectories in both cases. The time reversed trajectories according to the two parities, $x_{-}$($x_{+})$ for odd (even) parity, of the active noise are

\begin{figure*}[!htb]
\includegraphics[width=1\textwidth]{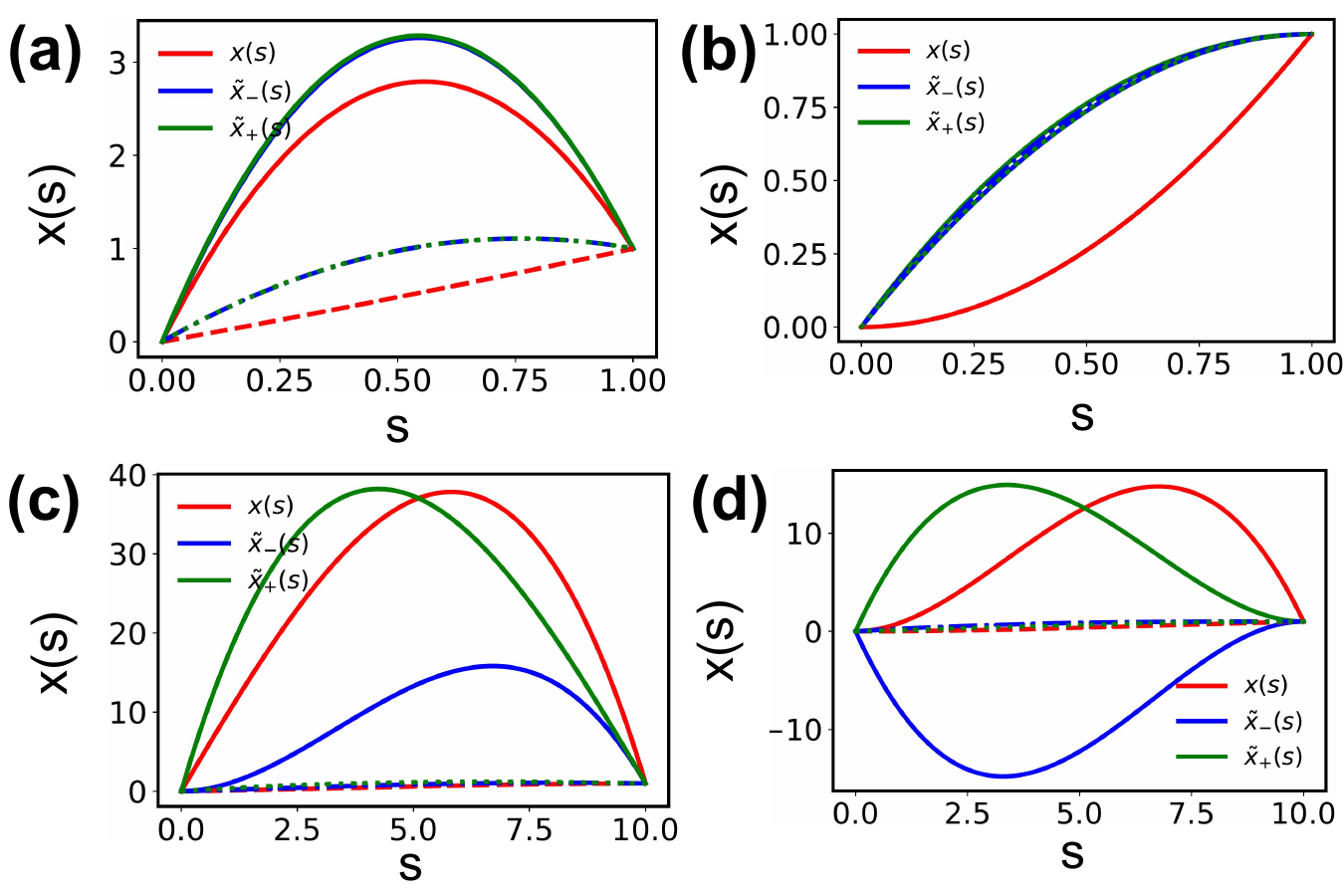}
\caption{ The time reversed trajectories in Eq \eqref{eq:timereversal} corresponding to the two parities for $\tau = 10$ and $C_1 = 10$. Trajectories in (a) and (c) do not pass through the fixed point with $\eta(0) = 0$ and in (b) and (d) pass through the fixed point. $t = 1$ in (a) and (b) and $t = 10$ in (c) and (d).   }
\label{fig:timereversal}
\end{figure*}

\noindent\begin{minipage}{0.5\linewidth}
\begin{align*}
&-\dot{\tilde{x}}_{-}(s)+\sqrt{2D_a}\tilde{\eta}_{-}(s) = C_1 \nonumber\\
 & \tau^2\ddot{\tilde{\eta}}_{-}(s) =-\tilde{\eta}_{-}(s)-\frac{\sqrt{2D_a}}{2D}C_1 \nonumber\\
\end{align*}
\end{minipage}
\begin{minipage}{0.5\linewidth}
\begin{align}
& -\dot{\tilde{x}}_{+}(s)-\sqrt{2D_a}\tilde{\eta}_{+}(s) = C_1 \nonumber\\
 & \tau^2\ddot{\tilde{\eta}}_{+}(s) =\tilde{\eta}_{+}(s)-\frac{\sqrt{2D_a}}{2D}C_1
 \label{eq:timereversal}
\end{align}
\end{minipage}
The plot of the time reversed trajectories under odd and even parities are shown in Fig \ref{fig:timereversal}. When $t < \tau$ the both time reversed trajectories are same regardless of whether they are passing through the fixed point (Fig \ref{fig:timereversal}-(a) and Fig \ref{fig:timereversal}-(b)). They are very different when $t\approx \tau$ as seen in Fig \ref{fig:timereversal} -(c) and Fig \ref{fig:timereversal}-(d).  

\begin{figure*}[!htb]
\includegraphics[width=1\textwidth]{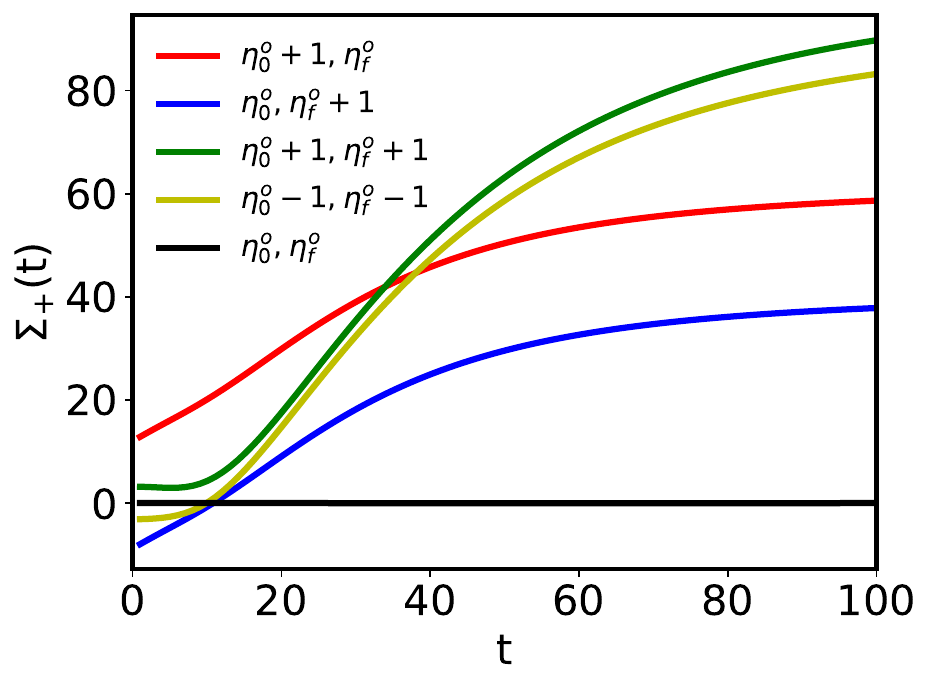}
\caption{The entropy production along the MPP (black) for $\tau = 100$ with optimal noises $(\eta_0^{(o)},\eta_f^{(o)})$ as in Fig \ref{fig:mpp} and the extremum paths with suboptimal noises. }
\label{fig:entropy}
\end{figure*}

In the odd parity case, the entropy production from Eq \eqref{eq:entropyproduction} reads \cite{dabelow2019irreversibility}
\begin{equation}
\Sigma_{-} = \int_0^tds\biggl[\frac{1}{D}\mathbf{f}(s).(\dot{\xb}(s)-\sqrt{2D_a}\etab(s))-2\tau\dot{\etab}(s).\etab(s) \biggr].
\end{equation}
 In particular for a free active particle in one dimension we get
\begin{equation}
\Sigma_{-}(t) =-\int_0^tds2\tau\dot{\eta}(s)\eta(s) = \tau (\eta_0^2-\eta_f^2).
\label{eq:sigmanegative}
\end{equation}
We recover the passive trajectory in Eq \eqref{eq:passive} in the large $\tau$ limit of the MPP in Eq \eqref{eq:largetau} when $\eta_0\approx 0$ and $\eta_f\approx 0$. Eq \eqref{eq:sigmanegative} too shows that the entropy production is zero along the MPP. Using $\langle\dot{\eta}(s)\eta(s)\rangle = -\frac{1}{2\tau^2}$ \cite{dabelow2019irreversibility}, we get $\langle\Sigma_{-}(t)\rangle = \frac{t}{\tau}$.   

In case of even parity, the entropy production from Eq \eqref{eq:entropyproduction} becomes
\begin{equation}
\Sigma_{+} = \int_0^tds\biggl[\frac{1}{D}\dot{\xb}(s).(\mathbf{f}(s)+\sqrt{2D_a}\etab(s))-2\tau\dot{\etab}(s).\etab(s) \biggr],
\end{equation}
which for free active particle in one dimension is
\begin{equation}
\Sigma_{+} = \int_0^tds\biggl[\frac{1}{D}\dot{x}(s)\sqrt{2D_a}\eta(s)-2\tau\dot{\eta}(s)\eta(s) \biggr]. 
\label{eq:sigmapositive}
\end{equation}

Entropy production is a measure of how far the system is from equilibrium. In the long time limit $t\rightarrow \infty$, the entropy production from Eq \eqref{eq:sigmapositive}  along the extremum trajectories in Eq \eqref{eq:trajectory} saturates as seen in Fig \ref{fig:entropy}. Similar to the behavior of $\Sigma_{-}$ in Eq \eqref{eq:sigmanegative} vanishing along the MPP, $\Sigma_{+}$ also vanishes along the MPP as seen in Fig \ref{fig:entropy}-(a). This is because the optimal noise $(\eta^{(o)}_0,\eta^{(o)}_f)$ is small at large $\tau$ (Fig \ref{fig:solution_space}) in which case we recover the passive particle in a thermal bath limit as discussed in Sec \ref{sec: aoup}. Thus for large $\tau$, the system behaves as an equilibrium system for which entropy production is zero. Larger the deviation from the optimal noise, larger is the entropy production thus driving the system away from equilibrium. The average entropy production $\langle\Sigma_{+}(t)\rangle = \left(1+\frac{D_a}{D}\right)\frac{t}{\tau}$ as derived in Appendix \ref{appendix: average entropy production} increases linearly with time very different from the saturation behavior of $\Sigma_{+}$ which is a result of fixed end points.

\section{Conclusion}
\label{sec:conclusion}
We have studied the MPP for AOUP active systems. OM formalism has been used to obtain the transition probabilities of these systems. The extremum of the OM integral is then obtained using the variational principle similar to the Lagrange equation in classical mechanics to obtain the extremum path of the AOUP particles. Out of all the extremum paths, the MPPs are the ones that has the minimum OM integral value. 

Free AOUP dynamics is analytically solvable. The extremum paths are solved for fixed start and end points of the trajectories, $x_0$ and $x_f$, and fixed start and end OU noise $\eta_0$ and $\eta_f$. We explored the phase space of the extremum paths by varying the persistence time $\tau$ and the noise $\eta_0$ and $\eta_f$ keeping the diffusion constants $D$ and $D_a$ constant. Unlike a passive particle which moves in a straight line from $x_0$ and $x_f$, the AOUP particle follows a curved path which approaches the passive trajectory at large $\tau$. The AOUPs with higher swim velocities deviates more from the straight line trajectories. The AOUP has an unstable fixed point which depends only on the active noise and particles passing through the fixed point slows down resulting in less curved trajectories. The entropy production along the extremum trajectories saturates to a maximum value, unlike the average noise that increases linearly with time. The active noise can be odd or even under time reversal and accordingly the trajectories of the AOUP are different. We also looked at the differences in entropy production along both types of trajectories. This work would be useful to control the trajectories of artificial active colloidal systems.  

The author acknowledges financial support by DST-SERB, India through the Startup Research Grant SRG/2022/000598 and MATRICS Grant MTR/2022/000281. The author also acknowledges financial support from the Additional Competitive Research Grant (PLN/AD/2022-23/3) from Birla Institute of Technology and Science, Pilani. The author also thanks DST-FIST for the computational resources provided to the Department of Physics, BITS Pilani.

\appendix

\section{Derivation of the Transition probability}
\label{appendix: onsagermaclup}
 The combined variable $\yb = (\xb,\etab)$ for $N$ particles, follows the Langevin equation by combining Eq \eqref{langevin} and $\tau\dot{\etab}_i(t)=-\etab_i(t)+\xib_i(t)$
\begin{equation}
\dot{y}_{i\alpha} = F_{i\alpha} + \xi_{i\alpha},
\end{equation}
with $F_i = (\mathbf{v}_i+\sqrt{2D_a}\etab_i,-\frac{1}{\tau}\etab_i)$. The Gaussian white noise $\langle\xi_{i\alpha}\rangle = 0$ and $\langle\xi_{i\alpha}(t)\xi_{j\beta}(0)\rangle = 2D_{i\alpha,j\beta}\delta(t)$, where $D_{i\alpha,j\beta} = 0$ is the diffusion tensor.
Following Ref \cite{yasuda2022most, risken1984solutions} we obtain for the path probability of a given trajectory $\yb(s)$ starting at $\yb_0$ and ending at $\yb(t)$ as
\begin{equation}
p(\yb(t)\vert\yb_0)  \propto \exp\biggl[-\int_0^t\frac{ds}{4}\left(F_{i\alpha}(\yb(s))-\dot{y}_{i\alpha}\right)D^{-1}_{i\alpha j\beta}\left(F_{j\beta}(\yb(s))-\dot{y}_{j\beta}\right)\biggr].
\end{equation}
For free AOUP particle the diffusion tensor is
\begin{equation}
D_{i\alpha j\beta} = \delta_{ij}
\begin{pmatrix}
D & 0 \\
0 & 1 \\
\end{pmatrix},
\end{equation}
which gives Eq \eqref{onsagermaclup} in the text.

\section{Exact solutions of Eq \eqref{eq:extremumpath}}
\label{app:solution}

The solution for the extremum path and the corresponding noise of Eq \eqref{eq:extremumpath} that satisfies the boundary conditions read
\begin{align}
x(s) & = \csch (t/(2 \tau))^2 \biggl[D_a \tau ((-
x_0+x_f) \cosh (s/\tau)+(x_0-x_f) \cosh ((s-t)/\tau)+(x_0+x_f)\times\nonumber \\ & (-1+\cosh (t/\tau)))+ D\sqrt{2D_a} \tau (-\eta_f s+\eta_f t-s \eta_0-\eta_f t \cosh (s/\tau)+t \eta_0 \cosh ((s-t)/\tau)+ \nonumber\\ 
 &(\eta_f s+s \eta_0-t \eta_0) \cosh (t/\tau))+D (-t x_0+s (x_0-x_f)) \sinh (t/\tau)+D_a (-t x_0+s (x_0-x_f)) \times\nonumber\\ &\sinh (t/\tau)+\sqrt{2} D_a^{3/2} \tau (-\eta_f s+\eta_f t-s \eta_0-\eta_f t \cosh (s/\tau)+t \eta_0 \cosh ((s-t)/\tau)+(\eta_f-\eta_0) \tau\times\nonumber\\& \sinh (s/\tau)+\cosh (t/\tau) (\eta_f s+s \eta_0-t \eta_0+(-\eta_f+\eta_0) \tau \sinh (s/\tau))+2 (\eta_f-\eta_0) \tau \sinh (s/(2 \tau))^2\times\nonumber \\& \sinh (t/\tau))\biggr]/\left[4 D_a \tau-2 (D+D_a) t \coth (t/(2 \tau))\right],
\label{eq:trajectory}
\end{align}
while for the noise we get
\begin{align}
\eta(s) & = e^{t/\tau} \biggl[2 D_a (\eta_f+\eta_0) \tau+2 D_a (\eta_f-\eta_0) \tau \cosh((s-t)/\tau)-2 D_a (\eta_f+\eta_0) \tau \cosh(t/\tau)+\nonumber\\& 2 (D+D_a) t (\eta_f-\eta_0 \cosh(t/\tau)) \sinh(s/\tau)- \sqrt{2D_a} (x_0-x_f) (-\sinh(s/\tau)+\sinh((s-t)/\tau)+\nonumber\\&\sinh(t/\tau))+2 \cosh(s/\tau) (D_a (-\eta_f+\eta_0) \tau+(D+D_a) t \eta_0 \sinh(t/\tau))\biggr]/\biggl[-(D+D_a)t-2 D_a \tau+\nonumber\\&4 D_a e^{t/\tau} \tau+e^{(2 t)/\tau} ((D+D_a) t-2 D_a \tau)\biggr]
\label{eq:noise}
\end{align}

\section{The average entropy production along a trajectory}
\label{appendix: average entropy production}
Applying the expression of the Langevin equation in Eq \eqref{langevin} in 1D for a single particle  to Eq \eqref{eq:entropyactivebath} we get
\begin{align}
\langle\Sigma\rangle & = \int_0^tds\frac{1}{D}\langle\left(\dot{x}(s).\sqrt{2D_a}\eta(s)-2\tau\dot{\eta}(s)\eta(s)\right)\rangle  \nonumber \\
& = \int_0^tds\frac{1}{\tau}\left(\frac{D_a}{D}+1\right) \nonumber\\ & = \frac{t}{\tau}\left(\frac{D_a}{D}+1\right). 
\label{eq:entropyactivebath}
\end{align}
Here we have used $\xi$ and $\eta$ in Eq \eqref{langevin} are uncorrelated, and $\langle\eta(s)\eta(s^{\prime}\rangle = \frac{1}{2\tau}\exp(-\vert s-s^{\prime}\vert/\tau)$.

\bibliography{main}
\end{document}